\documentclass[aps,prl,amssymb,showpacs,twocolumn]{revtex4-1}

\newcommand{\sigmab}{\mbox{\boldmath $\sigma $}}
\newcommand{\taub}{\mbox{\boldmath $\tau $}}

\newcommand{\bA}{{\bf A}}

\newcommand{\Nmath}{\mathcal{N}}
\newcommand{\Qmath}{\mathcal{Q}}

\newcommand{\bq}{{\bf q}}

\newcommand{\bP}{{\bf P}}
\newcommand{\bS}{{\bf S}}
\newcommand{\bD}{{\bf D}}

\newcommand{\br}{{\bf r}}

\newcommand{\rT}{{\rm T}}

\newcommand{\ua}{\uparrow}
\newcommand{\da}{\downarrow}

\newcommand{\beq}{\begin{equation}}
\newcommand{\beqn}{\begin{eqnarray}}
\newcommand{\eeq}{\end{equation}}
\newcommand{\eeqn}{\end{eqnarray}}
\newcommand{\nn}{\nonumber}

\newcommand{\eap}{easy-axis pspin}
\newcommand{\epp}{easy-plane pspin}

\usepackage{graphics,graphicx,amsmath}
\usepackage{amssymb}
\usepackage{amsthm}

\usepackage{tabularx,float}
\usepackage{epsfig}

\usepackage{subfigure}

\usepackage{bm}
\usepackage{wasysym}

\usepackage{bm}
\usepackage{color}
\usepackage{verbatim}

\def\tit#1#2#3#4#5{{#1}{\bf #2}, #3 (#4)}

\begin{document}



\author{Y. Lian$^1$, A. Rosch$^2$, and M. O. Goerbig$^1$}
\affiliation{
\centerline{$^1$Laboratoire de Physique des Solides, Univ. Paris-Sud, Universit\'e Paris-Saclay, CNRS, UMR 8502, F-91405 Orsay Cedex, France}
}
\affiliation{
\centerline{$^2$Institut f\"ur Theoretische Physik, Universit\"at zu K\"oln, D-50937 Cologne, Germany}
}

\title{SU(4) Skyrmions in the $\nu=\pm 1$ Quantum Hall State of Graphene}

\begin{abstract}

We explore different skyrmion types in the lowest Landau level of graphene at a filling factor $\nu=\pm 1$. 
In addition to the formation of spin and valley pseudospin skyrmions, we show that another type of spin-valley entangled skyrmions can be stabilized in graphene
due to an approximate SU(4) spin-valley symmetry that is affected by sublattice symmetry-breaking terms.
These skyrmions have a clear signature in spin-resolved density measurements on the lattice scale, and we discuss
the expected patterns for the different skyrmion types.

\end{abstract}

\pacs{73.43.Lp, 73.21.-b, 81.05.Uw}

\maketitle

Originally proposed in the framework of nuclear physics \cite{skyrme}, skyrmions have found physical reality in condensed-matter systems as 
topological textures of two-dimensional (2D) ferromagnets (FMs). Probably its conceptually purest form is realized in 2D electrons in a strong magnetic 
field $B$ \cite{sondhi} -- since their kinetic energy is quenched into highly degenerate Landau levels (LLs), all electrons spontaneously align their spins 
to minimize their Coulomb energy when there are as many electrons in a single LL as flux quanta threading the system. In the lowest LL, 
this corresponds to a filling factor $\nu=1$, whereas in graphene the same situation is encountered also at $\nu=-1$, due to particle-hole 
symmetry \cite{antonioRev,goerbigRev}. In both systems, skyrmions carry an electric charge given by their winding and have a lower energy than 
simple spin-flip excitations. In GaAs heterostructures, skyrmion formation yields a rapid decay of the magnetization in the vicinity of $\nu=1$, as measured in 
NMR experiments \cite{barretNMR}. 
More recently, skyrmions have regained interest \cite{Rosch1} after their discovery in chiral magnets \cite{skyrmExp1} and thin magnetic layers on 
heavy-metal substrates \cite{skyrmExp2}. They are promising candidates for spintronics application as they can easily be manipulated by ultrasmall 
currents \cite{Rosch2}. 
While characterized by the same type of winding numbers, skyrmions in these materials differ from quantum Hall skyrmions since 
they are bosonic quasiparticles and do not carry a quantized charge \cite{Rosch3}.

\begin{figure}
\centering
\includegraphics[width=\columnwidth ,angle=0]{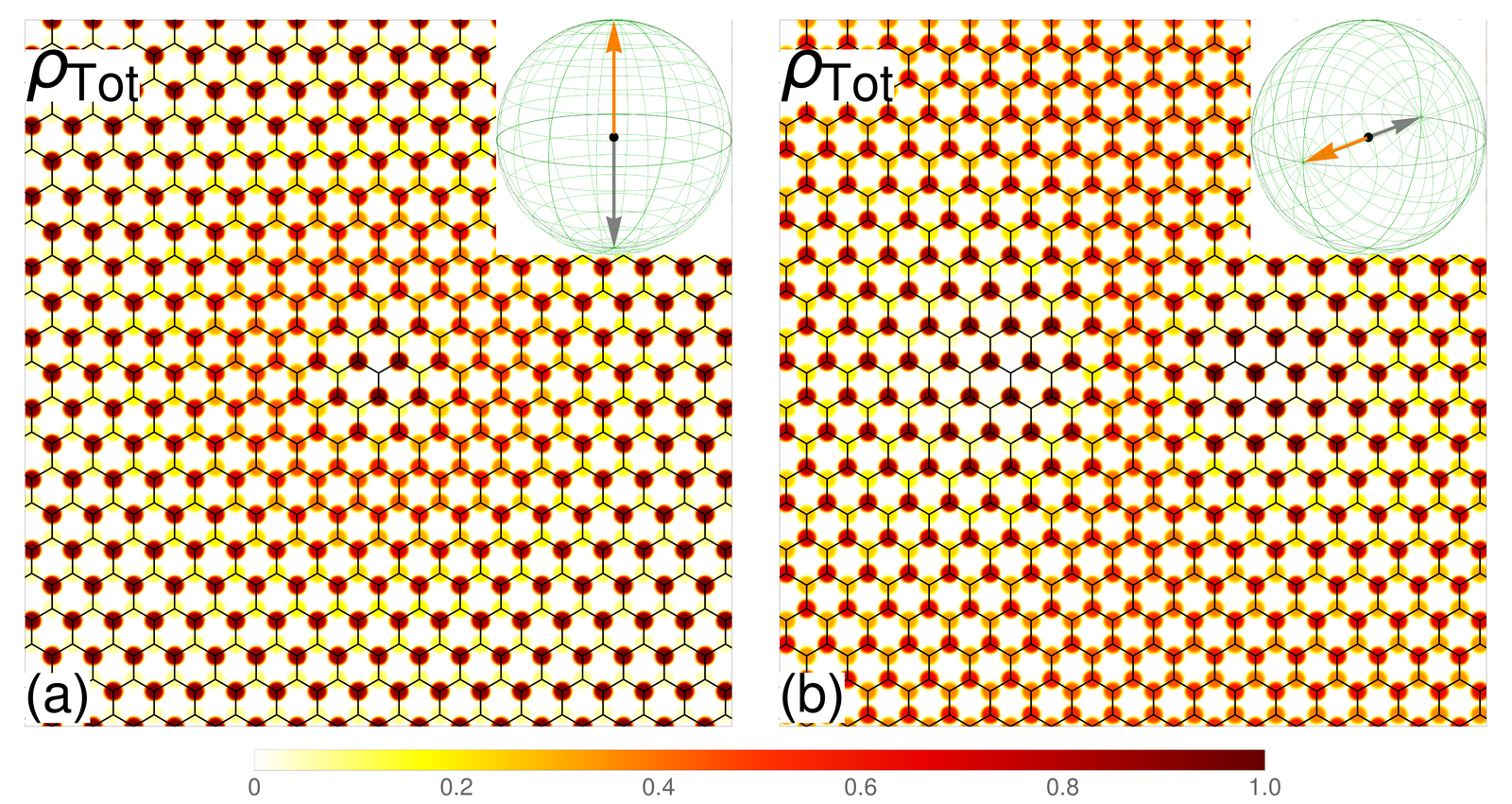} 
\caption{\footnotesize{Pseudospin skyrmions in an easy-axis [(a)] and an \epp\ [(b)] FM background. The two subfigures sketch the 
lattice-resolved total electronic density $\rho_{\rm tot}$ in the $n=0$ LL. 
The insets show the pseudospin Bloch sphere, which is entirely covered. In each Bloch sphere, the orange and gray arrows show the polarization of the FM background at $r\rightarrow \infty$ and in the center, respectively. At intermediate values of $r$, all other parts of the Bloch sphere are explored. The spin (not shown) is homogeneously polarized. 
For illustration, we have used much smaller skyrmion sizes (on the order of some lattice spacings) than encountered in an experimental situation (see text). (a) and (b) correspond to point 3 and 2 in Fig.~\ref{fig:PDS} respectively.
}}
\label{fig:EAPskyrm}
\end{figure}

A promising material that combines the conceptual simplicity of quantum Hall (QH) systems and the direct accessibility as a surface material is graphene. 
Moreover, graphene is characterized by an additional pseudospin (pspin) reflecting the two relevant valleys for its low-energy electronic 
properties \cite{antonioRev}. Since the Coulomb energy respects to great accuracy this pspin symmetry \cite{goerbigRev}, one encounters a particular form 
of \textit{SU(4) ferromagnetism} \cite{goerbig,yang,doucot} that allows for a much richer variety of skyrmions involving the valley pspin. Although valley
skyrmions have been studied in other materials \cite{shayegan}, the identity between valley and sublattice in the central $n=0$ LL \cite{goerbigRev} 
makes graphene an ideal candidate for a direct measurement of valley skyrmions, e.g. within spin-resolved scanning tunneling spectroscopy (STS).
Because all electrons of a particular valley thus reside on a single sublattice,
the valley pspin can be directly visualized by the sublattice occupation. This
would allow for a direct measurement of pspin skyrmions that are, e.g., depicted in Fig.~\ref{fig:EAPskyrm} and 
their size as a function of $B$.

In this Letter, we illustrate the different skyrmion types in graphene at $\nu=\pm 1$.
Beyond the expected spin and pspin skyrmions, we find a phase diagram with highly unusual skyrmions with spin and pspin entanglement. 
Whereas such skyrmions naturally arise in a purely SU(4)- \cite{yang,doucot} or more generally in any SU($K$)-symmetric model \cite{lillie}, 
their occurence in symmetry-broken situations has remained an open issue. Apart from a theoretical classification of the 
different skyrmion types in terms of Bloch spheres, we show how these skyrmions can be identified by their spin- and lattice-resolved electronic 
densities. Such densities are precisely accessible in STS, and our results may therefore be a guide in the 
spectroscopic identification of the different skyrmions in graphene, beyond spin skyrmions in the abovementioned other systems.

Our study is based on the non-linear sigma model
\beq\label{eq:model}
E[Z(\br)] = E_{\rm SU(4)}[Z(\br)] + E_{\rm sb}[Z(\br)]
\eeq
in terms of the spatially varying CP$^3$ field \cite{yang,lillie,skyrmBook,girvin} 
\beq
Z(\br) =\left[\upsilon_{K\uparrow}(\br),\upsilon_{K\downarrow}(\br),\upsilon_{K'\uparrow}(\br),\upsilon_{K'\downarrow}(\br)\right]^{T}
\eeq
whose four complex components 
represent the spin and pspin amplitudes in $n=0$, and $\br=(x,y)$ is the planar coordinate. 
Its first term is SU(4)-symmetric, 
\beqn\label{eq:Sym}
E_{\rm SU(4)}[Z(\br)] &=& 2\rho_{\rm S}\int d^2r \bD Z^{\dagger}(\br)\cdot \bD Z(\br)\\
\nn
&& + \frac{1}{2} \int d^2r d^2r' \rho_{\rm topo}(\br) V(\br-\br')\rho_{\rm topo}(\br'),
\eeqn
with the spin stiffness $\rho_{\rm S}=e^2/16\sqrt{2\pi}\epsilon l_B$ \cite{sondhi,moon}, the gradient 
$\bD Z = \nabla Z(\br) - [Z^{\dagger}(\br)\nabla Z(\br)] Z(\br)$, and the magnetic length $l_B=\sqrt{\hbar/eB}$.
The second term is the Coulomb interaction $V(\br)=e^2/\epsilon |\br|$
between the charge-density fluctuations that are, at $\nu=\pm 1$, identical to the topological charge density
$\rho_{\rm topo}(\br)= -(i/2\pi)[\bD Z(\br)^{\dagger} \times \bD Z(\br)]_z$ 
\cite{moon,EzawaCP3}. Apart from the SU(4)-symmetric term, the model also hosts 
symmetry-breaking terms,
\beq\label{eq:SB}
E_{\rm sb}[Z(\br)]= \frac{\Delta_Z}{2}\int \frac{d^2r}{2\pi l_B^2} \left[u_{\perp}\left( P_x^2 + P_y^2\right) + u_z P_z^2 - S_z\right],
\eeq
with the spin and pspin magnetizations
\beq
\bS=Z^{\dagger}(\br)(1\otimes \sigmab)Z(\br), \;  \bP=Z^{\dagger}(\br)(\sigmab\otimes 1)Z(\br),
\eeq
respectively, where $\sigmab=(\sigma_x,\sigma_y,\sigma_z)$ combines the three Pauli matrices. 
(An explicit expression of the spin and pspin densities, in terms of the CP$^3$-field components can be found in the Supplementary Material \cite{SM}.)
The parameters $u_z$ and $u_\perp$, which are presented in units of the Zeeman energy $\Delta_Z$,
describe, e.g., the pspin-symmetry breaking 
due to out-of-plane \cite{fuchs} or inplane \cite{inplane_phon} lattice distortions, or a symmetry breaking of the interaction at the lattice scale
\cite{goerbig}, and have been estimated to be all on the order of $0.1...0.2$ meV$\times B{\rm [T]}$, while the Zeeman effect is in the 
same range $\Delta_Z\simeq 0.1$ meV$\times B{\rm [T]}$. For realistic magnetic fields, this is much smaller than the leading (interaction) energy scale 
$\rho_S\sim e^2/\epsilon l_B\simeq 50$ meV$\times \sqrt{B{\rm [T]}}/\epsilon$. However, we emphasize that, while the hierarchy of energy scales is
well corroborated, the precise values of $u_z$ and $u_\perp$ are unknown and are likely to depend on the substrate. We therefore use them
as phenomenological parameters in our study.

\begin{figure}
\centering
\includegraphics[width=0.8\columnwidth, angle=0]{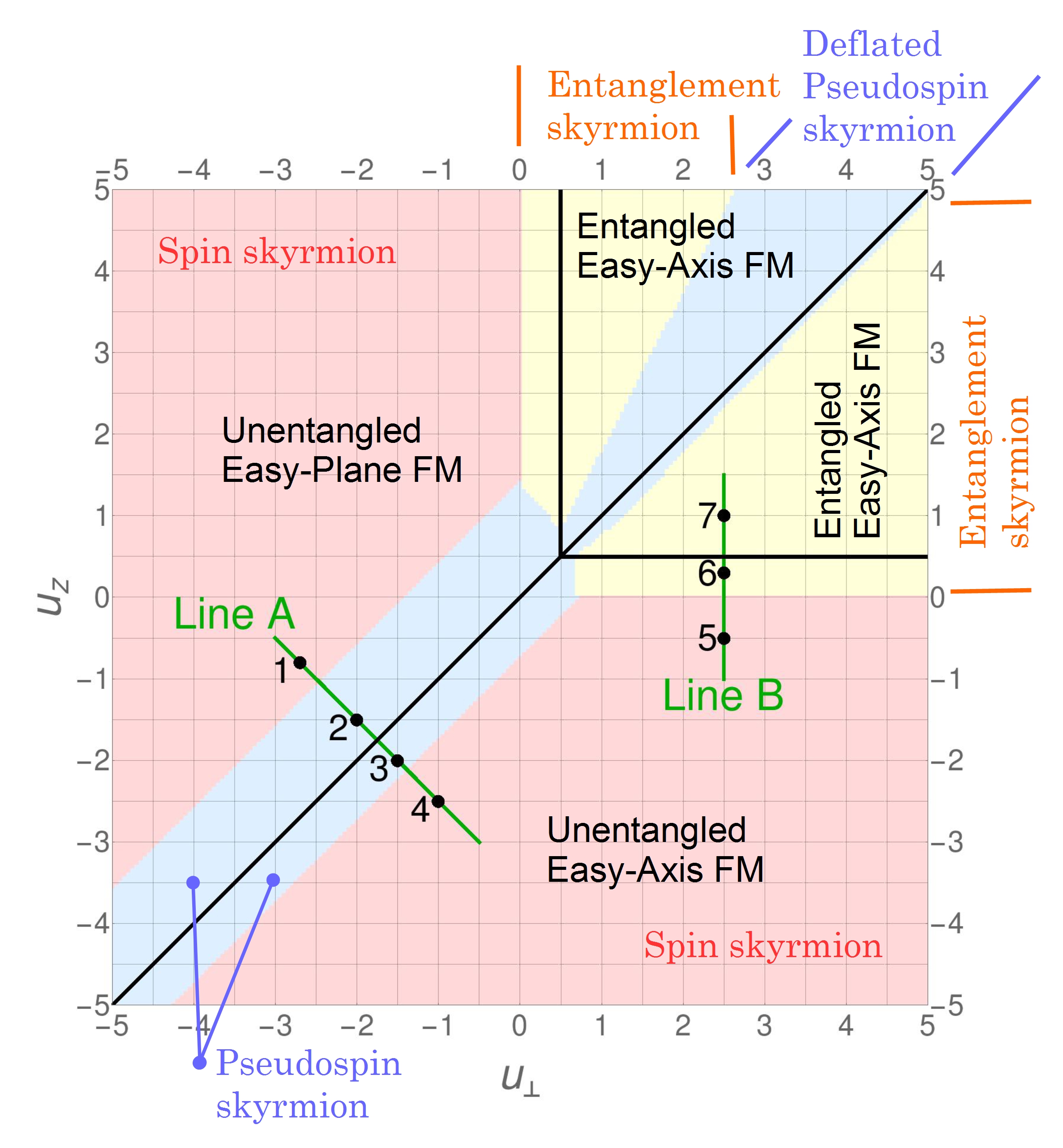}
\caption{\footnotesize{Phase diagram of $\Qmath=1$ skyrmions (labelled in red/blue/yellow) and the pspin FM background state (sketched in black), 
as a function of $u_z$ and $u_\perp$ (in units of the Zeeman energy $\Delta_Z$)
The parameters used correspond to a field $B=10$\,T.
}}
\label{fig:PDS}
\end{figure}

Because at large distances from their center, skyrmions approach the underlying FM background state, let us first discuss the phase diagram 
of homogeneous FM states described by a normalized spinor $Z(\br)=F$, 
similarly to Refs.~\cite{herbut} and \cite{kharitonov} at $\nu=0$. 
These states minimize the leading SU(4)-symmetric energy functional (\ref{eq:Sym}), $E_{\rm SU(4)}=0$, since all gradient terms vanish, and the 
symmetry-breaking terms (\ref{eq:SB}) thus determine the FM phase diagram (Fig.~\ref{fig:PDS}). Since the Zeeman term acts solely on the spin, and 
spin and pspin magnetic orders coexist at $\nu=\pm1$, all phases display a homogeneous spin magnetization in the $z$-direction. For $u_z\leq 1/2$ 
or $u_\perp\leq 1/2$, the spin and pspin magnetizations are disentangled, and one obtains an \epp\ FM, with e.g. $F=(1,0,\pm 1,0)^\rT/\sqrt{2}$,
for $u_z>u_\perp$ and an \eap\ FM, with $F=(1,0,0,0)^\rT$ or $F=(0,0,1,0)^\rT$, for $u_z < u_\perp$, 
in addition to a full spin polarization. We stress that in $n=0$ valley and sublattice are identical in the sense that the wave functions of an electron in 
a specific valley have only components on a particular sublattice \cite{SM}. 
The \eap\ FM therefore takes the form of a charge-density wave
with all spin-polarized electrons localized on a single sublattice, whereas both sublattices are equally populated in an \epp\ FM. 

The most interesting phases are obtained for $u_z,u_\perp>1/2$ where the spin and pspin magnetizations are partially entangled due to energetic frustration. 
According to Eq.~(\ref{eq:SB}), the pspin contribution to the anisotropy energy is lowered when all components of the pspin magnetization are minimized. As 
an extreme case, we consider a superposition $F=(1,0,0,1)/\sqrt{2}$ with spin-up electrons on the $A$-sublattice and spin-down particles on the $B$-sublattice, 
such that $\bP=0$. 
Somewhat counterintuitively, this state whose spin-density pattern is \textit{antiferromagnetic} remains a particular \textit{SU(4) FM} since it can be 
obtained from a pure spin (and pspin) FM via a rotation in the  SU(4) space. 
The drawback of such state with $\bP=0$ is a cost in the spin contribution (i.e. Zeeman energy) to the anisotropy energy in Eq.~(\ref{eq:SB}), because the amplitude of the spin magnetization $|\bS|$ also vanishes according to the equation $|\bS|=|\bP|=|\cos\alpha|$ that is valid \cite{doucot} for a generic CP$^3$-spinor. 
Therefore, this state can only be realized in the limit $\Delta_Z\rightarrow 0$.  
For finite $\Delta_Z$, $u_z>1/2$ and $u_{\perp}>1/2$, 
energy optimization leads to states with {\em partially} entangled spin and pspin, with either easy-plane ($u_z>u_\perp$) or easy-axis 
character ($u_z < u_\perp$).

To compute the phase diagram of CP$^3$ skyrmions with topological charge $\Qmath=1$,
we use that the by far largest contribution to the skyrmion energy is given by the gradient term in Eq.~(\ref{eq:Sym}). 
Minimizing this contribution, we obtain a skyrmion with energy $E_{\rm sk}=4\pi\rho_{\rm S}$ of the form
\beq\label{eq:skyrm}
Z_{\rm skyr}(x,y)=\Nmath(r)^{-1}\left[ (x+iy) F - \lambda(r) C \right],
\eeq
with constant $\lambda(r)=\lambda_0$ and  $r=|\br|$. $F$ is the CP$^3$ spinor of the FM background described above, and $\Nmath(r)$ ensures 
the normalization of $Z_{\rm skyr}(\br)$.
Due to the SU(4) symmetry and scale invariance of the gradient term, $C$ can be chosen as an arbitary spinor perpendicular to $F$, and also 
the size of the skyrmion obtained from $\lambda_0$ is not fixed.
These parameters are fixed by the remaining, much smaller anisotropy terms and the Coulomb energy. While the anisotropy terms favor small skyrmions, 
the Coulomb energy increases their size. For a quantitative analysis,  one has to take into account that for a constant $\lambda(r)=\lambda_0$  
the slow $1/r$ decay of the idealized SU(4) skyrmion causes a logarithmic divergence of the anisotropy energies Eq.~(\ref{eq:SB}). 
To obtain the asymptotically exact skyrmion energetics \cite{sondhi} 
and to avoid this divergency, it is sufficient to parametrize $\lambda(r) = \lambda_0 \exp(-r^2/\kappa \lambda_0^2)$. For each value of $u_\perp$ 
and $u_z$, we therefore minimize $\delta E = E[Z_{\rm skyr}(\br)] - E[Z(\br)=F]$ using $\lambda_0, \kappa$ and four variational angles 
characterizing $C$ \cite{SM}.
Typical skyrmion sizes obtained from this optimization are on the order of 50...100 graphene lattice spacings for realistic parameters. 
Notice that this is much larger than shown in our figures, where we have used a smaller skyrmion size  that corresponds to unphysical magnetic 
fields ($B\sim 1000$ T). However, the patterns are simpler to visualize and can easily be upscaled to realistic sizes.

The resulting skyrmion phases are shown in Fig.~\ref{fig:PDS}. Let us first concentrate on the rather simple cases $u_z\leq 0$ or $u_\perp\leq 0$, 
where the background $F$ is a product state of a spin and a pspin FM.
Either a spin or a pspin skyrmion can be formed to accomodate the $\Qmath=1$ topological charge. Charge excitations of minimal energy are mostly spin skyrmions
\beq\label{eq:Sskyrm}
Z_{\rm spin} (x,y) = \Nmath(r)^{-1} \psi^{\rm P}\otimes \left[ (x+iy)\psi_\ua^{\rm S} - \lambda(r) \psi_\da^{\rm S} \right],
\eeq
where the spinors $\psi_\ua^{\rm S}=(1,0)^{T}$ and $\psi_\da^{\rm S}=(0,1)^{T}$ represent the spin orientation and $\psi^{\rm P}$ is the (homogeneous) pspin 
component, which is unaffected by a pure spin texture. Generally the spinors can be represented in terms of the the four angles $\theta_{\rm S}$, $\phi_{\rm S}$ 
and $\theta_{\rm P}$, $\phi_{\rm P}$ that describe the spin and pspin polarizations on their respective Bloch spheres, with
\beqn\label{eq:angles1}
\psi^{\rm I} & = & \left[\cos(\theta_{\rm I}/2), e^{i\phi_{\rm I}}\sin(\theta_{\rm I}/2)\right]^{\rm T},\nonumber\\
\chi^{\rm I} & = & \left[-e^{-i\phi_{\rm I}}\sin(\theta_{\rm I}/2), \cos(\theta_{\rm I}/2)\right]^{\rm T},
\eeqn
for $ {\rm I}={\rm S},{\rm P}$. The spinors $\psi_\ua^{\rm S}$ and $\psi_\da^{\rm S}$ in Eq. (\ref{eq:Sskyrm})
correspond then to $\psi^{\rm S}(\theta_{\rm S}=0)$ and 
$\chi^{\rm S}(\theta_{\rm S}=0)$ respectively.

At $u_z\sim u_\perp$ and $u_z, u_\perp\le1/2$, it becomes energetically favorable to form pspin instead of spin skyrmions, 
\beq
Z_{\rm pspin}(x,y) = \Nmath(r)^{-1} \left[ (x + iy)\psi^{\rm P} - \lambda(r) \chi^{\rm P} \right]\otimes \psi_\ua^{\rm S},
\eeq
where we have $\theta_{\rm P}=0$ and $\theta_{\rm P}=\pi/2$ in $\psi^{\rm P},\chi^{\rm P}$ for the \eap\ \epp\ FM background, respectively. 
The pspin skyrmion in an \eap\ FM background is represented as a wrapping of the Bloch sphere in the inset of Fig.~\ref{fig:EAPskyrm}(a), as well as 
in a lattice-resolved image Fig.~\ref{fig:EAPskyrm}(a), for a set of parameters corresponding to point 3 in Fig.~\ref{fig:PDS}. 
While the CP$^3$-fields $Z(\br)$ only provide an envelope function in a continuum description, the lattice-resolved patterns can be obtained by 
a convolution with Gaussian functions representing the atomic wave functions on the lattice sites \cite{SM}.
The electronic density is concentrated on the $A$ sublattice at the skyrmion center $r=0$, whereas solely the $B$ sublattice is populated at $r\rightarrow\infty$.
The situation is more involved for a pspin skyrmion in an \epp\ FM (Fig.~\ref{fig:EAPskyrm}(b), for parameters corresponding to point 2 in Fig.~\ref{fig:PDS}). 
Since the pspin polarization is bound to the $xy$-plane at $r=0$ (gray arrow) and at $r\rightarrow \infty$ (orange arrow), both sublattices are equally populated there. 
However, because the pspin polarization explores all points of the Bloch sphere, the south pole at some point $\br_1$ and the north pole at $\br_2=-\br_1$, 
this yields the double-core structure in the lattice-resolved density plot Fig.~\ref{fig:EAPskyrm}(b), 
where solely the $A$ ($B$) sublattice is populated at $\br_1$ ($\br_2$). 
This is reminiscent of bimerons in bilayer quantum Hall systems in GaAs heterostructures \cite{moon,girvin,bimeron}.

The predominance of pspin skyrmions at $u_z\sim u_\perp$ is a consequence of a partial symmetry restoration at the transition $u_z=u_\perp$ -- the pspin component 
in Eq.~(\ref{eq:SB}) is then proportional to $u_z\bP\cdot\bP$, and all pspin orientations are equally possible. A deformation of the pspin 
texture thus becomes very soft, accompanied by no energy cost, while the full spin polarization allows one to minimize the Zeeman energy in Eq.~(\ref{eq:SB}). 
Similarly to spin skyrmions with a vanishing Zeeman gap \cite{sondhi}, the size of the pspin skyrmion diverges, apart from a logarithmic correction, as 
$\lambda_0/l_B \sim (e^2/\epsilon l_B\Delta|u_z-u_\perp|)^{1/3}$,
when approaching $u_z=u_\perp$ along line A in Fig.~\ref{fig:PDS}, as one may understand from a simple scaling analysis of the competing terms: 
while the pspin symmetry-breaking in Eq.~(\ref{eq:SB}) scales as $\sim \lambda_0^2 |u_\perp - u_z|$, the Coulomb interaction in Eq.~(\ref{eq:Sym})  
scales as $\sim e^2/\epsilon \lambda_0$.

\begin{figure}
\centering
\includegraphics[width=\columnwidth ,angle=0]{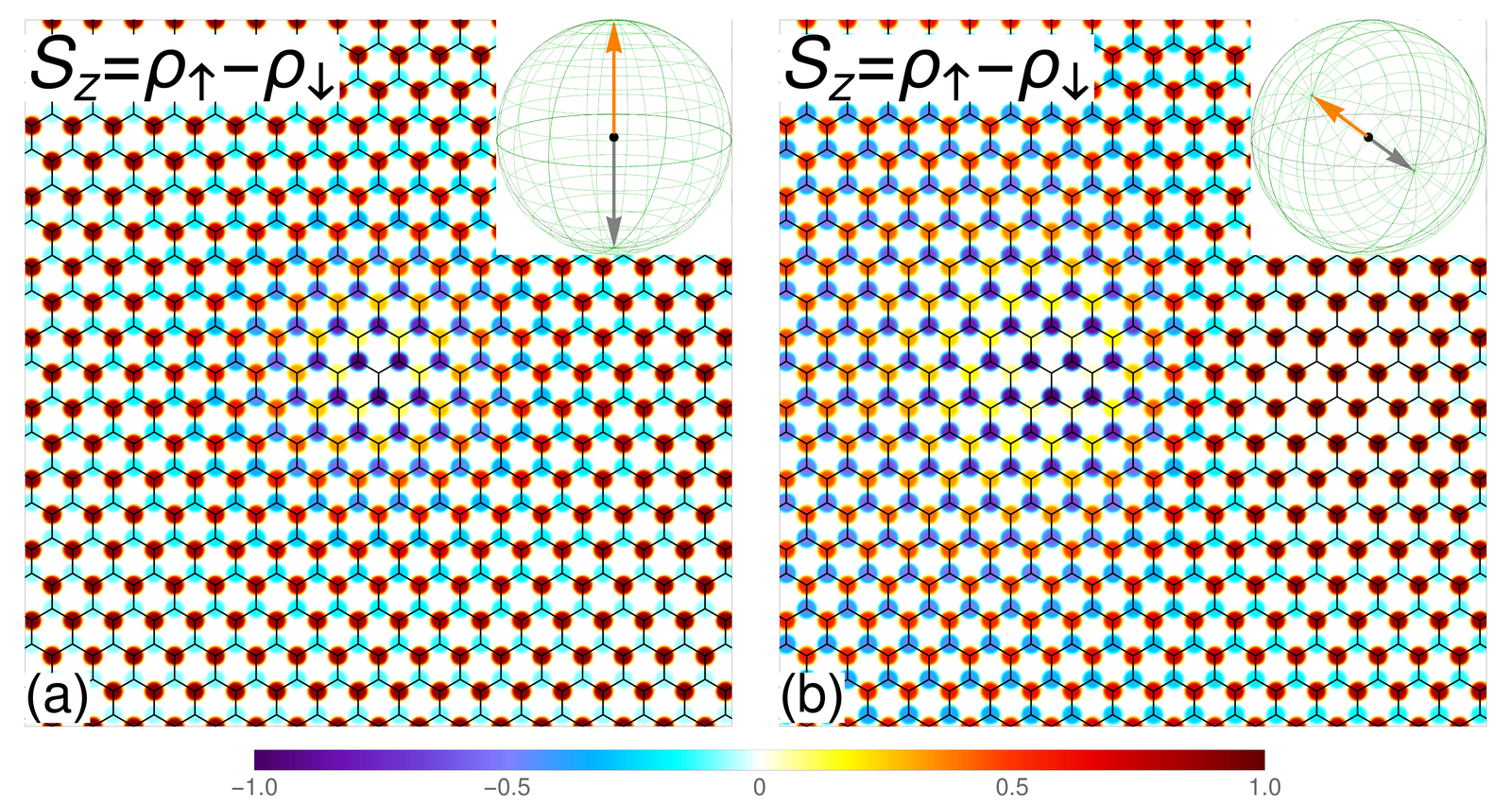}
\caption{\footnotesize{Entanglement skyrmions in an unentangled (a) and an entangled (b) FM background. 
The two subfigures sketch the lattice-resolved profiles of the $z$-component of the spin magnetization $S_{\rm z}$ in a color plot (blue for spin up and red for spin down densities). 
Insets show the entanglement Bloch sphere spanned by $\alpha$ and $\beta$ (see Eq.~\ref{eq:entAlpha}). In each Bloch sphere, the orange and gray arrows show the directions 
corresponding to the FM background and to the skyrmion center, respectively. (a) and (b) correspond to point 6 and 7 in Fig.~\ref{fig:PDS} respectively.
}}
\label{fig:EntSkyrm}
\end{figure}

The probably most exotic skyrmion types are obtained for $u_z,u_\perp\geq 0$ (yellow  in Fig.~\ref{fig:PDS}), where spin-pspin entanglement 
is energetically favored. A normalized CP$^3$ spinor is described by six angles. Whereas the first four  have been introduced in Eq.~(\ref{eq:angles1}), 
the remaining two ($\alpha$, $\beta$) can be viewed as angles on a \textit{third} Bloch sphere that describes the \textit{entanglement} 
between spin and pspin~\cite{doucot}
\beq \label{eq:entAlpha}
\Psi = \cos\frac{\alpha}{2}\, \psi^{\rm P}\otimes\psi^{\rm S} + e^{i\beta} \sin\frac{\alpha}{2}\, \chi^{\rm P}\otimes\chi^{\rm S}.
\eeq
This allows us to define the \textit{entanglement skyrmion} as an 
SU(4) texture that fully covers the third (entanglement) Bloch sphere, see the insets of Fig.~\ref{fig:EntSkyrm}, where the north and south 
poles correspond to no entanglement. This skyrmion can be formed both in an unentangled background [orange arrow pointing to the north pole in the 
inset of Fig.~\ref{fig:EntSkyrm}(a)] and in a FM background with non-zero entanglement -- in the latter case, the arrow representing the spinor $F$ points 
away from the poles [inset of Fig.~\ref{fig:EntSkyrm}(b)]. 

As we have pointed out above, the fingerprint of entanglement is a locally antiferromagnetic pattern,
and entanglement is thus better visible in a lattice-resolved plot of the spin magnetization rather than in plots of the different spin densities 
(such as in Fig. \ref{fig:EAPskyrm}). We therefore plot $S_z=\rho_\ua-\rho_\da$ in Fig.~\ref{fig:EntSkyrm} for the profile of entanglement skyrmions. 
(The separate patterns for $\rho_\ua$ and $\rho_\da$ for the same skyrmion types can be found in \cite{SM}.)  
Fig.~\ref{fig:EntSkyrm}(a) corresponds to the 
entanglement skyrmion in an unentangled FM background (point 6 in Fig.~\ref{fig:PDS}). We notice that also the skyrmion center is unentangled, with all
electrons on a single sublattice. In contrast to a pspin skyrmion in an \eap\ FM background, it is maximally entangled at $r\sim\lambda_0$, where 
one notices the abovementioned antiferromagnetic pattern, with all spin-up electrons on the B and all spin-down electrons on the A sublattice. 
Fig.~\ref{fig:EntSkyrm}(b), which represents the entanglement skyrmion in an {\em entangled} FM background (point 7 in Fig.~\ref{fig:PDS}), shows again a 
double-core structure. The two cores correspond to regions where $Z(r)$ has no entanglement, while the entangled FM background is manifest 
in the antiferromagnetic pattern. 
Finally we notice that one obtains again pspin skyrmions at $u_z\sim u_\perp$, $u_z, u_{\perp}>1/2$ (upper-right region in Fig.~\ref{fig:PDS}). However, 
to minimize the symmetry-breaking terms, they are partially entangled, i.e. the polarization explores regions of the entanglement 
Bloch sphere different from the poles. Hence, the modulus $|\cos\alpha|$ of the pspin polarization is decreased (``deflated pspin skyrmion'' in Fig.~\ref{fig:PDS}), 
and the density contrast would be reduced as compared to Fig.~\ref{fig:EAPskyrm}.

In conclusion, we have investigated different skyrmion types in graphene at $\nu=\pm 1$. Apart from the usual spin skyrmion, 
the valley pseudospin analogue yields distinctive charge patterns on the graphene lattice because valley and sublattice degrees of 
freedom are identical in the $n=0$ LL. Graphene is therefore an ideal system to probe skyrmions with a valley pspin texture. This can
in principle be achieved in lattice-resolved STS in the energy range corresponding to $n=0$. Since quantum-Hall skyrmions carry, 
in contrast to those in chiral magnets, electric charge, their density can be controlled by a back gate and one can thus achieve the
limit of few isolated skyrmions.
Most saliently, the large SU(4) symmetry of the leading terms in the non-linear sigma model yields exotic 
entanglement skyrmions stabilized for positive values of the parameters $u_z$ and $u_\perp$. These topological objects also have a clear 
fingerprint in the form of \textit{antiferromagnetic} patterns, e.g. in spin-resolved STS, even if they are manifestations of SU(4)-FM states. 
Our results show that using STS with a magnetic tip one can not only detect but also identify
the various skyrmion types and analyze their size as a function of $B$ \cite{sondhi,SM}. Notice that the relative weight of the parameters can to some extent
be tuned by the $B$-field and its orientation -- while the Zeeman energy depends on the total field, the pspin couplings only depend on its 
perpendicular component \cite{fuchs,inplane_phon}. If one, furthermore, combines magnetic tips with different orientations of the 
magnetization \cite{wiesendanger} one can actually map out locally 5 of the 6 angles parametrizing the CP$^3$ field [see Eq. (\ref{eq:entAlpha})]. 
Only the combination $\beta-\phi_P$ cannot be measure directly.
While we have concentrated the 
discussion on skyrmions with topological charge $\Qmath=1$, in topological sectors with higher charge the Coulomb repulsion is likely 
to break up a single charge $\Qmath$ into several charge-1 skyrmions that are eventually arranged into a lattice \cite{skyrmExp1,skyrmLattice}.

We acknowledge fruitful discussions with Markus Morgenstern. YL is funded by a scholarship from the China Scholarship Council.

\pagebreak

\setcounter{equation}{0}
\setcounter{figure}{0}

\renewcommand{\theequation}{S\arabic{equation}}
\renewcommand{\thefigure}{S\arabic{figure}}

\begin{widetext}

\section*{Supplementary Material}

\section{Sublattice occupation as pseudospin polarization}

This first section is meant to be a reminder of the intimite link between the valley index and the sublattice characteristic of the $n=0$ graphene Landau level,
based on a more detailed description in Ref. \onlinecite{goerbigRev}. This link finds its origin in the structure of the $n=0$ Landau-level wave functions,
which one obtains from a solution of the Hamiltonian \cite{goerbigRev}
\begin{equation}
{\cal H}_{\boldsymbol{q}}=\hbar v_{F}\sigma_{z}\otimes\boldsymbol{q}\cdot\taub ,
\end{equation}
where we introduced the four spinor representation
\beq\label{eq:1partSpin}
\Psi_{\boldsymbol{q}}=\left(\psi_{\boldsymbol{q}}^{A,K},\psi_{\boldsymbol{q}}^{B,K},\psi_{\boldsymbol{q}}^{B,K'},\psi_{\boldsymbol{q}}^{A,K'}\right)^{T} .
\eeq
Correspondingly in the Hamiltonian, $\sigma_z$ and $\taub=(\sigma_x,\sigma_y)$ corresponds to the valley and sublattice, respectively. Notice that, here, we
are interested only in the wave functions of the Landau-level problem, which do not depend on the spin degree of freedom. The latter would enter, formally, in
the one-particle Hamiltonian simply as an additional one-matrix that would yield an eight spinor in the form of two identical copies of the above four spinor. 
After the Peierls substitution $\hbar\bq\rightarrow\boldsymbol{\Pi}=\hbar\bq+e\bA(\br)$
and choosing the symmetric gauge, we obtain the eigenstates $\Psi_{n,m}^{\xi}$ for Landau orbit $m$ in Landau level $n$
\beqn\label{eq:LLspin}
\Psi_{n=0,m}^{\xi=K} & = & \left(0,\left|n=0,m\right\rangle ,0,0\right)^{T}\nonumber \\
\Psi_{n=0,m}^{\xi=K'} & = & \left(0,0,0,\left|n=0,m\right\rangle \right)^{T}\nonumber \\
\Psi_{n\neq0,m,\lambda}^{\xi=K} & = & \frac{1}{\sqrt{2}}\left(\left|n-1,m\right\rangle ,\lambda\left|n,m\right\rangle ,0,0\right)^{T}\nonumber \\
\Psi_{n\neq0,m,\lambda}^{\xi=K'} & = & \frac{1}{\sqrt{2}}\left(0,0,\left|n-1,m\right\rangle ,-\lambda\left|n,m\right\rangle \right)^{T}
\eeqn
and the Landau level energy 
\begin{equation}
E_{n,\lambda}=\lambda\frac{\hbar v_{F}}{l_{B}}\sqrt{2n}\,,
\end{equation}
where the band index $\lambda=\pm$ denotes the sign of the energy, and $\left|n,m\right\rangle$ is the same quantum-mechanical state as in the standard Landau 
quantization for electrons with parabolic energy dispersion -- $n$ denotes the Landau level and $m$ denotes the Landau orbit.

We notice that in the $n=0$ Landau level of monolayer graphene, the non-vanishing component for the eigenstate $\Psi_{n=0,m}^{\xi=K}$ 
and $\Psi_{n=0,m}^{\xi=K'}$ are $(B,K)$ and $(A,K')$ respectively, indicating that the sublattice A is empty for the eigenstate of 
$K$ valley, and sublattice B is empty for the eigenstate of $K'$ valley. In this sense, we can identify the sublattice
and the valley in the $n=0$ Landau level. 
We use the pseudospin to describe a generic superposition of the two eigenstates of $K$ and $K'$ valleys. 
Specifically, ``pseudospin up'' means an electron in the $K$ valley and occupies only the B sublattice, whereas a ``pseudospin down'' 
state refers to the valley $K'$ and the electron occupies only the A sublattice. Hence in the $n=0$ Landau level, the sublattice occupation 
unveils the pseudospin polarization, and a pseudospin texture state usually has distinguished patterns of sublattice occupations.

\section{CP$^{3}$-spinor and spin / pseudospin magnetization}

Eq. (5) in the main text can be expanded explicitly in the components of the CP$^3$-field $Z(\br)$. We denote its components by 
\begin{equation}
Z(\br)=\left[\upsilon_{K\uparrow}(\br),\upsilon_{K\downarrow}(\br),\upsilon_{K'\uparrow}(\br),\upsilon_{K'\downarrow}(\br)\right]^{T}.
\end{equation}
Notice that the sublattice index does no longer occur explicitly in the CP$^3$-field, and the spinor is therefore different from that (\ref{eq:1partSpin}) used 
above in the description of the one-particle quantum states. Indeed, it is redundant in the $n=0$ Landau level where it is identical to the valley pspin, whereas
it is fixed in all other Landau levels, as can be seen from the expressions (\ref{eq:LLspin}).
The $z$-component of the spin magnetization $S_{z}(\br)$ is 
\beq\label{eq:spinZ}
S_{z}(\br) = Z^{\dagger}(\br)(1\otimes\sigma_{z})Z(\br) = \left|\upsilon_{K\uparrow}(\br)\right|^{2}-\left|\upsilon_{K\downarrow}(\br)\right|^{2}+
\left|\upsilon_{K'\uparrow}(\br)\right|^{2}-\left|\upsilon_{K'\downarrow}(\br)\right|^{2},
\eeq
and the $z$-component of the pseudospin magnetization $P_{z}\left(r\right)$ is
\beq\label{eq:pspinZ}
P_{z}(\br) = Z^{\dagger}(\br)(\sigma_{z}\otimes1)Z(\br) = \left|\upsilon_{K\uparrow}(\br)\right|^{2}+\left|\upsilon_{K\downarrow}(\br)\right|^{2}-
\left|\upsilon_{K'\uparrow}(\br)\right|^{2}-\left|\upsilon_{K'\downarrow}(\br)\right|^{2}.
\eeq
The spin / pseudospin magnetization for the FM spinor $F$ can be expressed in its components by setting $Z(\br)=F$ in the above equations. 
For instance, the example $F=(1,0,0,1)^{T}/\sqrt{2}$, which corresponds to a fully spin-pseudospin entangled $SU(4)$ FM with antiferromagnetic 
pattern on thr lattice scale, gives $\bS=\bP={\bf 0}$.

The four components of the CP$^3$-skyrmion ansatz Eq. (6) can be written as 
\beq
\upsilon_{I}(\br) = \frac{w_{I}(\br)}{{\cal N}(\br)}, \;
w_{I}(\br) = \left(x+iy\right)F_{I}-\lambda(r)C_{I}
\eeq
where $I=K\uparrow,K\downarrow,K'\uparrow,K'\downarrow$ and the normalization factor is 
\beq
{\cal N}(\br)=\sqrt{\left|w_{K\uparrow}(\br)\right|^{2}+\left|w_{K\downarrow}(\br)\right|^{2}+\left|w_{K'\uparrow}(\br)\right|^{2}+
\left|w_{K'\downarrow}(\br)\right|^{2}}.
\eeq

Let us write down the components of the CP$^3$-field $Z_{\rm spin}(\br)$ {[}Eq. (6) in the main text{]} for a spin skyrmion embedded in 
the easy-axis FM background. The FM background spinor $F$ carries no entanglement and thus can be decomposed as $F=\psi^{P}\otimes\psi_{\uparrow}^{S}$ 
with $\psi^{P}=\left(1,0\right)^{T}$ and $\psi_{\uparrow}^{S}=\left(1,0\right)^{T}$.
According to Eq. (6) in the main text, the center spinor $C$ can also be decomposed as $C=\psi^{P}\otimes\psi_{\downarrow}^{S}$ 
with $\psi_{\downarrow}^{S}=\left(0,1\right)^{T}$. Therefore, the four components of $Z_{\rm spin}(\br)$ are
\beq
\upsilon_{K\ua}(\br) = \frac{x+iy}{x^{2}+y^{2}+\lambda^{2}(r)}, \;
\upsilon_{K\da}(\br) = \frac{-\lambda(r)}{x^{2}+y^{2}+\lambda^{2}(r)}, \;
\upsilon_{K'\ua}(\br) = \upsilon_{K'\da}(\br) = 0.
\eeq
The spin magnetization of the spin skyrmion $Z_{\rm spin}(\br)$ can be compared to the  $O(3)$ skyrmion \cite{skyrmBook}, described in terms of the magnetization
${\bf S}(\br)$.
Insertion of the above expressions into Eq. (5) of the main text yields
\beq
S_x(\br)=\frac{-2\lambda(r) x}{(\lambda(r)^{2}+x^{2}+y^{2})^{2}}, \;
S_y(\br)=\frac{2\lambda(r) y}{(\lambda(r)^{2}+x^{2}+y^{2})^{2}}, \;
S_z(\br)=\frac{x^{2}+y^{2}-\lambda(r)^{2}}{(\lambda(r)^{2}+x^{2}+y^{2})^{2}}.
\eeq
This is equivalent to the familiar form of $O(3)$ skyrmion \cite{skyrmBook} up to a global rotation of the spin texture along the $y$-direction of the spin 
magnetization space. One can verify that $\bS(\br)$ carries topological charge ${\cal Q}=1$.

\section{Minimization of $E\left[Z(\br)\right]$}
In the main text, we use the following non-linear sigma model 
\beqn\label{eq:model}
E[Z(\br)] & = & E_{\rm SU(4)}[Z(\br)] + E_{\rm sb}[Z(\br)]\\
E_{\rm SU(4)}[Z(\br)] & = & E_{\rm NLSM}[Z(\br)] + E_{\rm C}[Z(\br)] \\
E_{\rm NLSM}[Z(\br)] & = & 2\rho_{\rm S}\int d^2r \bD Z^{\dagger}(\br)\cdot \bD Z(\br) \\
E_{\rm C}[Z(\br)] & = & \frac{1}{2} \int d^2r d^2r' \rho_{\rm topo}(\br) V(\br-\br')\rho_{\rm topo}(\br')\\
E_{\rm sb}[Z(\br)] & = & \frac{\Delta_Z}{2}\int \frac{d^2r}{2\pi l_B^2} \left[u_{\perp}\left( P_x^2 + P_y^2\right) + u_z P_z^2 - S_z\right]
\eeqn
for the CP$^3$-field $Z(\br)$ to capture the ordering in the $N=0$ Landau level in monolayer graphene, at the particular filling factor $\nu=-1$. 
In the non-linear sigma model energy $E_{\rm NLSM}[Z(\br)]$, the spin stiffness is $\rho_{\rm S}=e^2/16\sqrt{2\pi}\epsilon l_B$, in terms of the 
magnetic length $l_B=\sqrt{\hbar/eB}$, and the gradient means $\bD Z = \nabla Z(\br) - [Z^{\dagger}(\br)\nabla Z(\br)] Z(\br)$. 
In the Coulomb energy $E_{\rm C}[Z(\br)]$, the Coulomb potential is $V(\br)=e^2/\epsilon |\br|$. At $\nu=\pm 1$, the excess charge density 
$\delta\rho_{\rm el}=\rho_{\rm el}-\rho_0$ is identical to the topological charge density
\beq
\rho_{\rm topo}(\br)= -(i/2\pi)[\bD Z(\br)^{\dagger} \times \bD Z(\br)]_z.
\eeq
In the symmetry-breaking energy $E_{\rm sb}[Z(\br)]$, the spin and pseudospin magnetizations are computed from the CP$^3$-field $Z(\br)$ as
\beq\label{eq:S_P}
\bS=Z^{\dagger}(\br)(1\otimes \sigmab)Z(\br), \;  \bP=Z^{\dagger}(\br)(\sigmab\otimes 1)Z(\br),
\eeq
or explicitly in components as mentioned in the previous section of this note.

The soliton solution of topological charge ${\cal Q}=1$ is described by the following ansatz [Eq. (5) in the main text]:
\beq\label{eq:skyrm}
Z_{\rm skyr}(x,y)=\Nmath(r)^{-1}\left[ (x+iy) F - \lambda(r) C \right],
\eeq
where $F$ and $C$ are (normalized) spinors for the FM background and the skyrmion center, respectively. They satisfy $F^{\dagger}C=0$ so that the origin 
of the $xy$-plane coincides with the skyrmion center. 
For a constant function $\lambda(r)=\lambda_0$ we have $E_{\rm NLSM}[Z_{\rm skyr}(\br)]=4\pi\rho_{\rm S}$, $E_{\rm C}[Z_{\rm skyr}(\br)]\sim \lambda_0^{-1}$ 
and $E_{\rm sb}[Z_{\rm skyr}(\br)]\sim \lambda_0^{2}$. For a generic monotonically decreasing function $\lambda(r)$, $E_{\rm NLSM}$ is slightly larger, 
but the scaling of $E_{\rm C}$ and $E_{\rm sb}$ remain the same.

The FM background spinor $F$ in $Z_{\rm skyr}$ can be determined by 6 angles according to Eq. (10) in the main text, whereas the center spinor $C$ 
needs only 4 angles because of the constraint $F^{\dagger}C=0$. We assume 
\beq
\lambda(r) = \lambda_0 \exp(-r^2/\kappa \lambda_0^2)
\eeq
to take radial deformation into account (explained in the main text). It contains two real parameters: $\lambda_0$ and $\kappa$. Each concrete CP$^3$-field 
$Z_{\rm skyr}$ describing a skyrmion is determined by $6+4+2=12$ real parameters. The energy functional $E[Z_{\rm skyr}]$ then becomes a function of the 12 parameters. 

Since $Z_{\rm skyr}(\br)$ can be understood as an interpolation between $F$ and $C$, i.e. a skyrmion is embedded in the FM background, the 6 angles in 
$F$ should be determined prior to the other parameters. This is achieved by minimizing $E_{\rm FM}=E[Z(\br)=F]=E_{\rm sb}[Z(\br)=F]$ for the spatially homogeneous
states. The result is presented 
in the main text in Fig.~2, where we draw black lines for the border between two regions of different types of FM background spinor.
In the next stage, for each pair of $(u_{\perp},u_z)$, we minimize the energy difference $E[Z_{\rm skyr}(\br)]-E_{\rm sb}[Z(\br)=F]$ with the FM 
background spinor $F$ in $Z_{\rm skyr}(\br)$ determined in the earlier step. This energy minimization gives the optimal values of the 4 angles in the center 
spinor $C$, and the two parameters $\lambda_0, \kappa$ in function $\lambda(r)$.

The phase diagram for ${\cal Q}=1$ skyrmions Fig.~2 in the main text is produced by performing the aforementioned two-stage energy minimization at 
each $(u_{\perp},u_z)$ point, and we then use the red, blue or yellow colors to label the type of skyrmion as the minimization result. 
The magnetic field $B$ is set to $10{\rm T}$ so that the ratio between the Zeeman energy $\Delta_Z$ and the Coulomb energy $e^2/\epsilon l_B$ is approximately $0.002$.

\section{Size of the skyrmions}

In the main text we discussed the divergence in the size of pspin skyrmion when $u_z\sim u_\perp$ and $u_z,u_\perp<1/2$. Here we compute the skyrmion size $R$ 
as an average of $r$ on the topological charge density $\rho_{topo}(\br)$
\beq
R=\int r\rho_{topo}(\br)d^{2}r ,
\eeq
and plot it in Fig.~\ref{fig:SkyrmSize} as a function of $u_z-u_{z0}$ along line A and line B in the phase diagram of ${\cal Q}=1$ skyrmion in the main text. 
Here, $u_{z0}$ denotes the value at the border between two regions of different FM background, where $u_z=u_\perp$ for line A and $u_z=1/2$ for line B. 
While we have used the wave function of the deformed skyrmion with $\lambda(r)$ given in the main text, its size is mainly governed by the bare size parameter $\lambda_0$ in the skyrmion ansatz. 
One would thus expect the same scaling and therefore should have same scaling behavior as for an undeformed skyrmion (with $\lambda(r)=\lambda_0$),
\beq\label{eq:scale}
\frac{\lambda_0}{l_B} \sim \left(\frac{e^2}{\epsilon l_B\Delta|u_z-u_\perp|}\right)^{1/3}.
\eeq
This is confirmed in Fig. \ref{fig:SkyrmSize}, where the size is fitted in a log-log plot in the inset (red lines). 
The numerically extracted exponent is indeed with $0.31$ very close to $1/3$, and we attribute the slight discrepancy to the deformation $\lambda(\br)$ 
of the skyrmions to render the symmetry-breaking terms non-divergent, as mentioned above. 
The divergence of the skyrmion size at the line $u_z\sim u_\perp$ therefore reflects the 
underlying transition line between
an easy-axis and an easy-plane pspin ferromagnetic background, as long as both are unentangled. 

The situation is different along line B cutting the transition line $u_z=1/2$ between an unentangled and an entangled \eap\ FM background, where 
the skyrmion size increases but does not diverge (blue line) since there is no evident symmetry restoration.
Indeed, the power law is cut off (see blue lines in the inset of Fig. \ref{fig:SkyrmSize}), and we use the fitting law
$\lambda \sim (|u_z-1/2| + C)^{-\gamma}$, with some constant $C$ describing the cutoff. The 
exponent $\gamma$ is again close to $1/3$, as expected from our simplified scaling analysis. Even if there is no fully developed divergence in the skyrmion
size, its increase unveils again a transition between different underlying FM background states.

\section{Visualization of a CP$^3$-skyrmion on honeycomb lattice}

In the main text, we visualize a CP$^3$-skyrmion $Z_{\rm skyr}(\br)$ by plotting the lattice-scale profiles of the electron density $\rho_{\rm Tot}(\br)$ 
and the $z$-component of the spin magnetization $S_{\rm z}(\br)$ for the CP$^3$-field $Z_{\rm skyr}(\br)$. The lattice-scale profiles are computed via 
Eq. (\ref{eq:S_P}), where the $\alpha=(\lambda,\sigma)$ components of the CP$^3$-field are convoluted with a form factor 
\beq
f_{\lambda}(\br)=\sum_{\br_{j}^{\lambda}}g(\br-\br_{j}^{\lambda})
\eeq
In the above equation, $\br_{j}^{\lambda}$ denotes the lattice vector of the $\lambda$-sublattice in the j'th unit-cell. The function $g(\br)$ 
represents the atomic wave function of the $p_z$-orbital and has been chosen to be Gaussian for illustration purpose. 
If we further neglect the overlap of $g(\br-\br_{j}^{\lambda})$ between atomic wave functions at different lattice sites, the expressions for 
the lattice-resolved total and spin densities read
\beqn
\rho_{\rm Tot}(\br) &= \sum_{\br_{j}^{\lambda}}\rho_{\rm Tot}(\br_{j}^{\lambda})\tilde{g}(\br-\br_{j}^{\lambda})\\
S_{\rm z}(\br) &= \sum_{\br_{j}^{\lambda}}S_{\rm z}(\br_{j}^{\lambda})\tilde{g}(\br-\br_{j}^{\lambda}),
\eeqn
respectively, where the occupation at sublattice A, B is given by $\rho_{\rm Tot}(\br_{j}^{A,B})$, and the $z$-component of spin magnetization at sublattice 
A, B is given by $S_{\rm z}(\br_{j}^{A,B})$. The latter are calculated similarly to Eqs. (\ref{eq:spinZ}) and (\ref{eq:pspinZ}) as
\beqn
\rho_{\rm Tot}(\br_{j}^{A,B}) &= Z^{\dagger}(\br_{j}^{A,B})\left[\frac{\sigma_{z}\pm1}{2}\otimes1\right]Z(\br_{j}^{A,B})\\
S_{\rm z}(\br_{j}^{A,B}) &= Z^{\dagger}(\br_{j}^{A,B})\left[\frac{\sigma_{z}\pm1}{2}\otimes\sigma_{z}\right]Z(\br_{j}^{A,B}),
\eeqn
where the $+$ sign in the projector is chosen for a site on the $A$-sublattice and $-$ for a site on the $B$-sublattice.
The function $\tilde{g}(\br)$ is again chosen to be Gaussian and normalized as $\tilde{g}(\br=0)=1$.

\section{Equivalent plots of Fig. 3}

In Fig. 3 of the main text, we show the lattice-resolved profiles of the spin magnetization $S_{z}(\br)=\rho_{\uparrow}(\br)-\rho_{\downarrow}(\br)$ 
for the entangled skyrmions. We have used pairs of high-contrast colors to stress the fact that the electron spin on different sublattices are of 
opposite directions. One can also plot the lattice-resolved profiles of the spin magnetization $\rho_{\ua}(\br)$ and $\rho_{\da}(\br)$ separately, 
as shown in Fig.~\ref{fig:Lattice-scale-profiles-pt6} and Fig.~\ref{fig:Lattice-scale-profiles-pt7}. Taking advantage of the fact that spin-up and 
spin-down electrons have non-vanishing amplitudes in different sublattices, Fig. 3 in the main text can be obtained by changing the color scheme of 
the $\rho_{\da}(\br)$ profile and combine it with the $\rho_{\ua}(\br)$ profile. Thus Fig.~\ref{fig:Lattice-scale-profiles-pt6} and 
Fig.~\ref{fig:Lattice-scale-profiles-pt7} in this note are equivalent to Fig. 3 in the main text.

\clearpage

\begin{figure}[p]
\centering
\includegraphics[width=0.6\columnwidth ,angle=0]{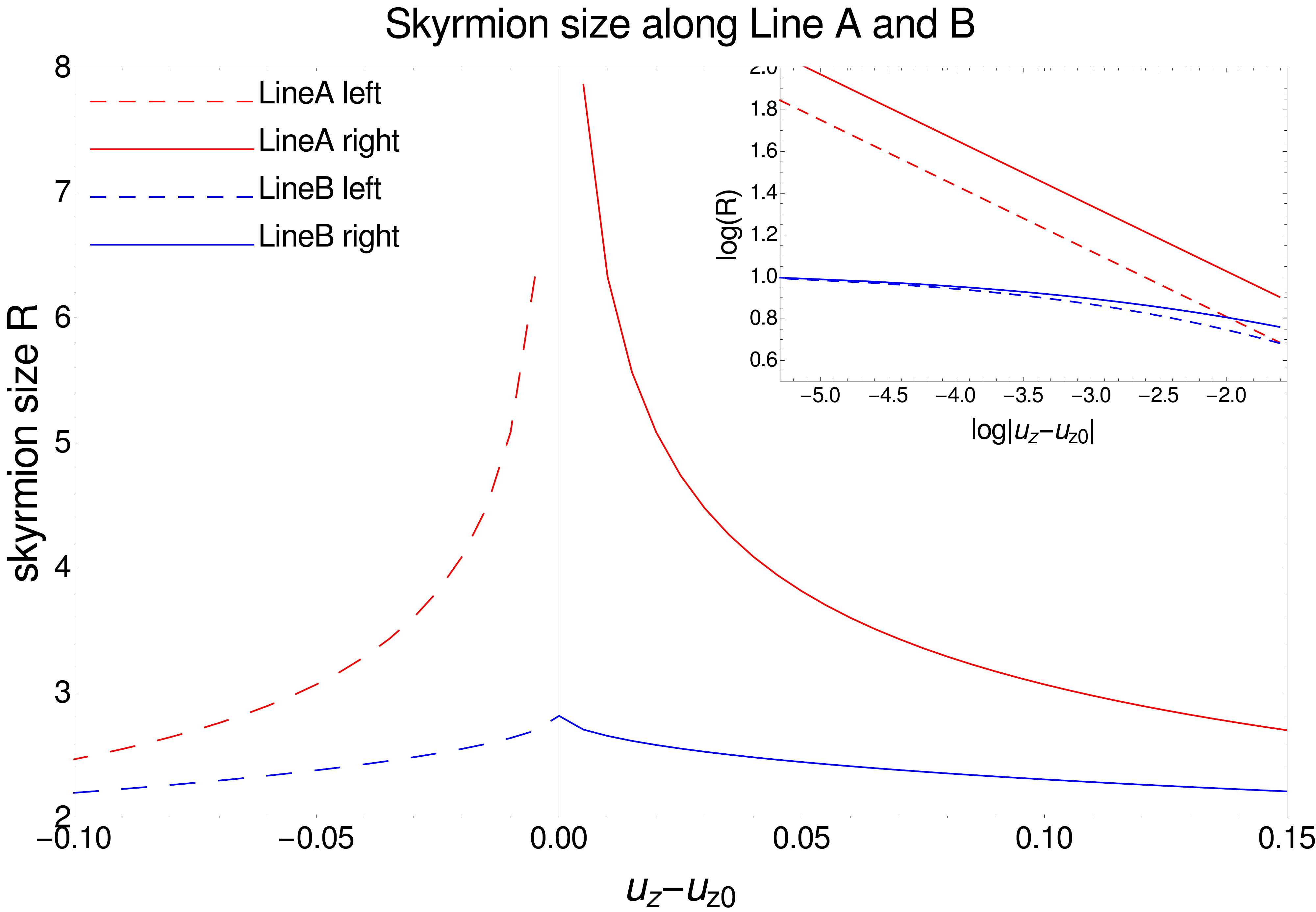}
\caption{\footnotesize{Skyrmion size (in unit of magnetic length $l_B$) as function of $u_z$, along line A and B in Fig.2 of the main text. }}
\label{fig:SkyrmSize}
\end{figure}

\clearpage

\begin{figure}[p]
\begin{centering}
\includegraphics[width=\columnwidth]{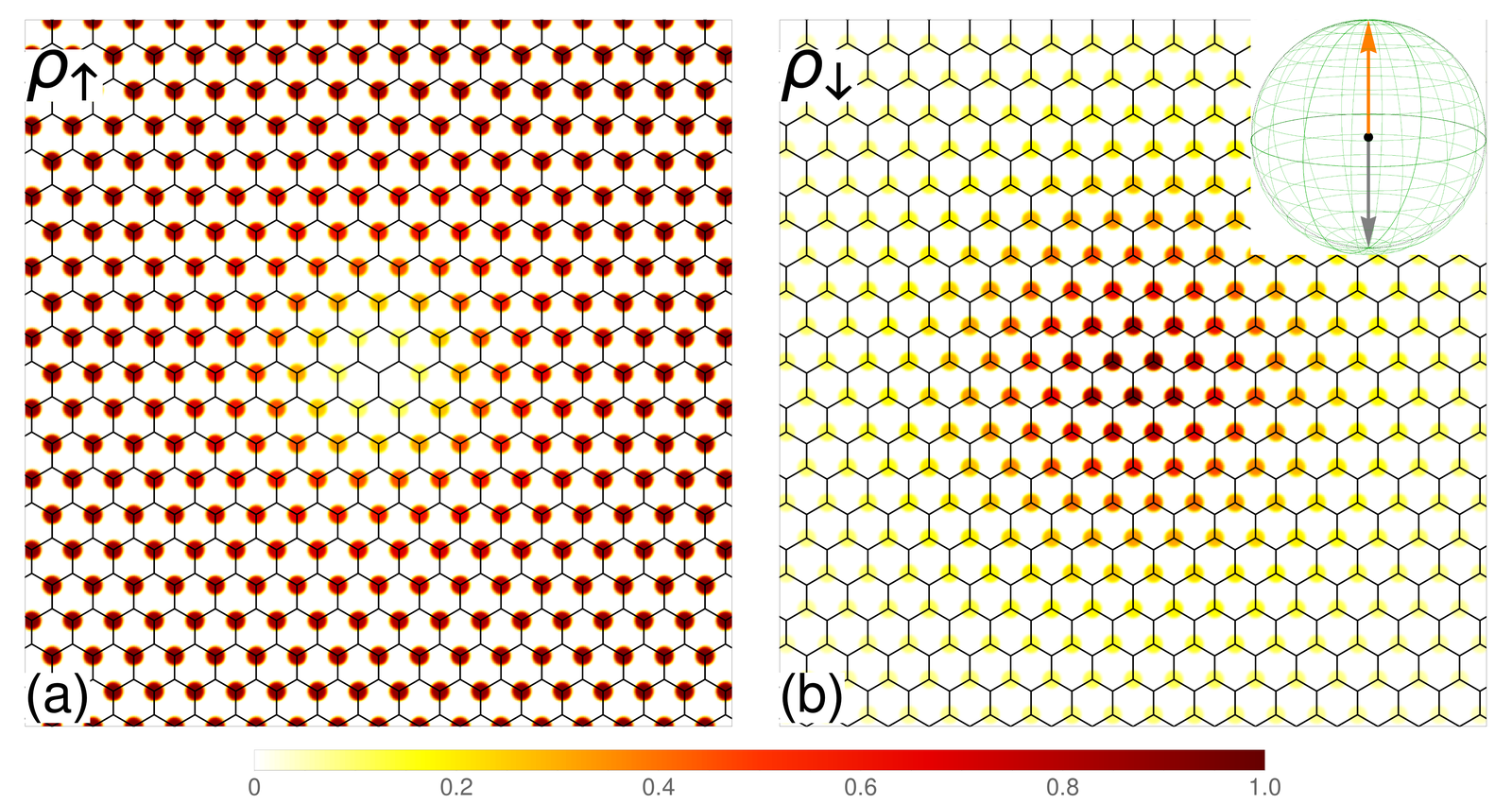}
\par\end{centering}
\caption{\label{fig:Lattice-scale-profiles-pt6} Lattice-scale profiles of
$\rho_{\uparrow}\left(r\right)$ {[}(a){]} and $\rho_{\downarrow}\left(r\right)$
{[}(b){]} for entanglement skyrmions embedded in the \textbf{unentangled}
easy-axis FM background. Such skyrmion appears as result of energy
minimization at point $6$ in Fig. 2 of the main text. }
\end{figure}
\begin{figure}[p]
\begin{centering}
\includegraphics[width=\columnwidth]{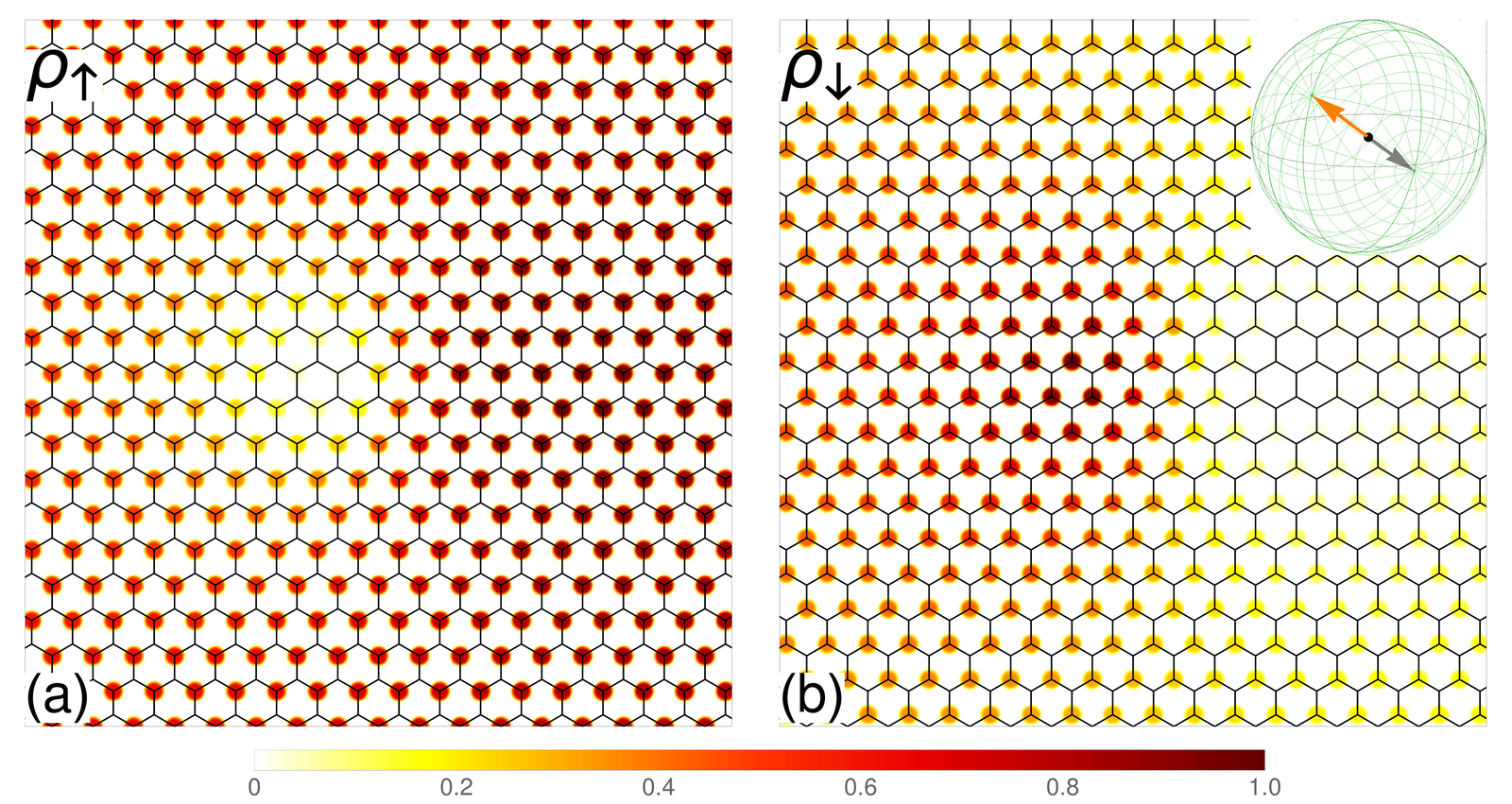}
\par\end{centering}
\caption{\label{fig:Lattice-scale-profiles-pt7} Lattice-scale profiles of
$\rho_{\uparrow}\left(r\right)$ {[}(a){]} and $\rho_{\downarrow}\left(r\right)$
{[}(b){]} for entanglement skyrmions embedded in the \textbf{entangled}
easy-axis FM background. Such skyrmion appears as result of energy
minimization at point $7$ in Fig. 2 of the main text. }
\end{figure}

\end{widetext}

\end{document}